\documentclass[%
  reprint,
  aps,
  prb,
  superscriptaddress,
  amsmath,amssymb,
]{revtex4-1}

\usepackage[utf8]{inputenc} 

\usepackage{graphicx} 

\usepackage[normalem,]{ulem} 
\usepackage[hyperref,table,dvipsnames,svgnames,]{xcolor} 
\usepackage{hyperref} 
\hypersetup{
  colorlinks = true,
  citecolor  = blue,
  linkcolor  = blue,
  urlcolor   = blue,
}

\usepackage{diagbox} 
\usepackage{dcolumn} 

\newcommand{\tableheader}[2]{\multicolumn{#1}{c}{#2}}

\usepackage{bm} 
\usepackage{siunitx} 

\newcommand{\units}[1]{\ensuremath{\,\,(#1)}}
\DeclareSIUnit \electron {e}

\newcommand{\E}{\ensuremath{\mathrm{e}}}
\newcommand{\PI}{\ensuremath{\mathrm{\pi}}}

\newcommand{\Del}[1]{\ensuremath{\mathinner{\mathrm{\Delta}#1}}}
\newcommand{\dif}[1]{\ensuremath{\mathinner{\mathrm{d}#1}}}

\renewcommand{\vec}[1]{\ensuremath{\bm{\mathrm{#1}}}}

\begin{document}
  \title{Ab Initio calculation of field emission from metal surfaces with atomic--scale defects}
  
  \author{H.~\surname{Toijala}}  
  \affiliation{Helsinki Institute of Physics and Department of Physics, University of Helsinki, PO Box 43 (Pietari Kalmin katu 2), 00014 Helsinki, Finland}
  
  \author{K. Eimre}
  \affiliation{Intelligent Materials and Systems Lab, Institute of Technology, University of Tartu, Nooruse 1, 50411 Tartu, Estonia}

  \author{A. Kyritsakis}%
  \email{andreas.kyritsakis@helsinki.fi; akyritsos1@gmail.com}
  \affiliation{Helsinki Institute of Physics and Department of Physics, University of Helsinki, PO Box 43 (Pietari Kalmin katu 2), 00014 Helsinki, Finland}

    \author{V. ~\surname{Zadin}}
\affiliation{Intelligent Materials and Systems Lab, Institute of Technology, University of Tartu, Nooruse 1, 50411 Tartu, Estonia}

  \author{F. ~\surname{Djurabekova}}
  \affiliation{Helsinki Institute of Physics and Department of Physics, University of Helsinki, PO Box 43 (Pietari Kalmin katu 2), 00014 Helsinki, Finland}
  \affiliation{National Research Nuclear University MEPhI, Kashirskoye sh. 31, 115409 Moscow, Russia}
  \date{\today}

  \begin{abstract}
  
    In this work we combine density functional theory and quantum transport calculations to study the influence of atomic--scale defects on the work function and field emission characteristics of metal surfaces. 
    We develop a general methodology for the calculation of the field emitted current density from nano-featured surfaces, which is then used to study specific defects on a Cu(111) surface.
    Our results show that the inclusion of a defect can significantly locally enhance the field emitted current density. 
    However, this increase is attributed solely to the decrease of the work function due to the defect, with the effective field enhancement being minute.
    Finally, the Fowler--Nordheim equation is found to be valid when the modified value for the work function is used, with only an approximately constant factor separating the computed currents from those predicted by the Fowler--Nordheim equation.
    
  \end{abstract}
  
  \maketitle
  
  \section{Introduction} \label{section:intro}
  
  Field electron emission (FE) from metal surfaces plays a crucial role in various aspects of modern technology.
  It is widely exploited in various devices that require cold electron sources \cite{deheer1995carbon,saito2000field,okano1996low,jensen1999field}, such as electron microscopes \cite{crewe1338visibility}, flat displays \cite{liu2012new,kuznetzov2010electron}, electron beam lithography \cite{vieu2000electron}, electrospray propulsion \cite{lemmer2017propulsion}, etc \cite{egorov2017field}.
  On the other hand, it often appears parasitically (in this case, it is usually referred to as dark current), limiting the performance or even igniting catastrophic vacuum breakdown events in a wide range of devices that involve metal surfaces exposed to high electric fields.
  Such devices span from micro- and nano-electronic capacitors \cite{lyon2013gap, ducharme2009inside}, macroscopic high-voltage devices such as vacuum interrupters and X-ray tubes \cite{latham1995,slade2007}, up to large-scale apparatuses such as fusion reactors \cite{juttner2001cathode,McCracken1980} and particle accelerators \cite{dobert2005high, kim2008proposal, clic, clic2016, nanni2015terahertz}.
  
  Vacuum breakdowns (also known as vacuum arcs) are initiated at sites on the metal surface which exhibit strong field electron emission \citep{latham1995, Anders}, even when the applied surface electric fields are in the range \numrange{30}{250} \SI{}{\mega\volt\per\meter}, far lower than the FE regime (\numrange{1}{10} \SI{}{\giga\volt\per\meter}).
  The systematic experimental study of such emitting sites \cite{beukema1972conditioning, Descoeudres2009_dc, Descoeudres2009} has shown an apparent field enhancement factor $\beta$ in the range \numrange{20}{140}. 
  In other words, the current-voltage characteristics of such sites are consistent with a local field $\beta$ times higher than the applied macroscopic one.
  The most common interpretation of this apparent field enhancement is geometric, i.e. explained by assuming the existence of microscopic protrusions on the metal surface that locally enhance the electric field \cite{beukema1972conditioning, Descoeudres2009_dc, Descoeudres2009}.
  However, $\beta$ values of the aforementioned order correspond to protrusions with an aspect ratio in the same range, which have never been observed on the studied surfaces, even when high-resolution electron microscopy is utilized \cite{Kildemo2004}.
  
  The above discrepancy indicates that the classical Fowler--Nordheim (FN) theory \cite{FN1928,Nordheim1928,MurphyG} with its standard interpretation for analyzing experimental data \cite{forbes2009use} might be insufficient to describe FE under certain conditions.
  The FN equation \cite{MurphyG} that canonically describes FE was derived assuming a mathematically flat planar metal surface with a given work function, a free--electron model for the metal bulk and a semi--classical approximation for the tunneling probability.
  It has already been shown that the FN equation is valid only for surfaces with radii of curvature larger than $\approx \SI{20}{\nano\meter}$ \cite{Cutler1993,CutlerAPL}.
  Although the behaviour of non-planar emitters has been studied extensively \cite{Zuber2002, EdgcombeDeduction, CabreraPRB, KXnonfn, KXGTF,GETELECpaper}, these works relied on analytical approximations for the surface potential barrier, the assumption of a smooth, well defined surface for the emitter and a semi-classical approach for the transmission probability.
  
  One mechanism that could possibly contribute in the explanation of the above unnaturally high apparent field enhancement factors is the local increase of the field emission current due to atomic--level surface defects such as adatoms, adsorbates, atomic steps etc.
  Such defects have already been shown to have a significant effect on the work function~\cite{Djurabekova2013}.
  Furthermore, field electron emission from atomic--scale features cannot be conceptually treated with the classical FN theory, since most of its assumptions are not valid.
  However, \textit{ab initio} computational techniques \cite{neugebauer1992adsorbate, lepetit2017electronic, Groth2014} offer the ability to overcome the above problems and directly calculate the emission current densities for any surface configuration, without adopting most of the mathematical simplifications of the classical theory.
  
  In this paper, we develop an \textit{ab initio} computational method to calculate the field emitted current from defective metal surfaces.
  Our method combines density functional theory (DFT) electronic structure calculations with quantum transport methods.
  We use our method to obtain the field emission characteristics from defective Cu surfaces, with three different types of features, namely adatoms, mono-atomic steps and two-layer pyramids.
  We find that the current density and the corresponding apparent field enhancement increase on the defective surfaces, compared to the clean one.
  However, this increase is attributed mainly to the decrease of the corresponding work function.
  The methodology developed here can be used in the future to perform similar calculations for various kinds of surface defects such as adsorbates, oxide layers and larger nano-features.

    The structure of the paper is the following:
    Section~\ref{sec:theory} will introduce the theoretical background for this work.
    Section~\ref{sec:methods} will describe the calculation of the field emission current.
    Section~\ref{sec:results} will show the results of the calculations and their interpretation using Fowler--Nordheim plots.
    Limitations of the methodology and the interpretation of the results will be discussed in section~\ref{sec:discussion}.
    Finally, the conclusions drawn from the work will be given in section~\ref{sec:conclusions}.
  
  \section{Theory} \label{sec:theory}
    \subsection{Field emission theory} \label{sec:theory-FN}
      Field emission is still most often described using the Fowler--Nordheim equation developed in the 1920s~\cite{Nordheim1927, FN1928, Nordheim1928}.
      The Fowler--Nordheim theory makes several approximations in order for the field emission current to be analytically solvable.
      The metal is assumed to be described by the free--electron jellium model, with the ionic potential having a discontinuity at the surface, where it changes to the vacuum potential.
      The surface is assumed to be perfectly flat, reducing the problem to one dimension.
      The jellium potential is smoothed by the image potential, which has its image plane at the jellium surface.

      Murphy and Good~\cite{MurphyG} give a rigorous mathematical treatment of the problem with the assumptions above, giving the emission current as an integral
      \begin{equation}
        J(\phi, F, T) = e \int_{0}^{\infty} \dif{E_{z}} N(E_{z}, T) D(E_{z}, \phi, F) \,, \label{eq:MG-J}
      \end{equation}
      where $E_{z} = \hbar^2 k_{z}^2 / (2 m)$ is the normal (to the emitting surface) energy component, $\vec{k} = (k_{x}, k_{y}, k_{z})$ is the wave vector, $m$ is the electron rest mass, $\phi$ is the work function, $F$ is the applied electric field on the surface and $T$ is the temperature.
      $N(E_{z}, T)$ is the supply function, which gives the normal energy--dependent flux of electrons to the surface, and $D(E_{z}, \phi, F)$ is the transmission probability, the probability for an electron to tunnel through the surface potential barrier into the vacuum.
      Using the above assumptions along with the semi-classical JWKB approximation \cite{landau} for the transmission coefficient, the integral can be solved analytically for low temperatures.
      For $T = 0$ it yields the Fowler--Nordheim equation
      \begin{equation} \label{eq:FNeq}
        J(\phi, F, T) = \frac{a F^2}{\phi \tau^2(F)} \exp\left[- \frac{\nu(F) b \phi^{3/2}}{F} \right] \,,
      \end{equation}
      where $a$ and $b$ are universal constants and $\nu(F)$ and $\tau(F)$ are well-known mathematical functions.
      Murphy and Good also give a correction factor to the Fowler--Nordheim equation for low non-zero temperatures.

      Forbes~\cite{Forbes2004, Forbes2007, *Forbes2007corr} gives interpretations of $\nu$ and $\tau$, linking them to the Gamow exponent $G = - \ln D(E_{z}, \phi, F)$ and its derivative with respect to $E_z$ at the Fermi level.
      The barrier enhancement factor
      \begin{equation}
        \nu(F) = \frac{G(E_{\text{F}}, \phi, F)}{G_{\text{ET}}(E_{\text{F}}, \phi, F)}
      \end{equation}
      is the ratio of the Gamow exponent at the Fermi level to that of the corresponding exact triangular barrier, while $\tau(F)$ is the same for the derivative of the Gamow exponent with respect to the normal energy.

      Two additional effects are often added to eq. \eqref{eq:FNeq} to obtain the technically complete Fowler--Nordheim equation~\cite{ForbesBarrier}.
      Due to geometric effects not taken into account when assuming the planar surface, the electric field at the surface can be larger than the macroscopic applied electric field $F$.
      This is taken into account using the field enhancement factor $\beta$, replacing $F$ with $\beta F$.
      Also, a multiplicative constant $\lambda$ is added to the equation to include the effects of the band structure and other effects of real materials which do not appear in the jellium model.
      With these changes, the emission current density is given by \cite{ForbesBarrier}
      \begin{equation} \label{eq:FN_comp}
        J(\phi, F, T) = \lambda \frac{a \beta^2 F^2}{\phi \tau^2(\beta F)} \exp\left[- \frac{\nu(\beta F) b \phi^{3/2}}{\beta F} \right] \,.
      \end{equation}
    \subsection{Calculation from first principles} \label{sec:theory-changes}
      \subsubsection{Deviation from Fowler--Nordheim theory} \label{sec:theory-changes-general}
        In the current work, we abandon simplifying approximations used in the standard FN theory, in order to provide a more realistic description of the field emitting system from first principles.

        First, the supply function $N(E_z, T)$ is modified to allow for arbitrary density of states.
        In the standard FN theory which is based on the free electron model, the electron states are considered to be plane waves, with an isotropic parabolic dispersion relation $E = \hbar^2 k^2 / 2m$ and a density of states that has a square root dependence on the total kinetic energy.
        In this work, we use the density of states as calculated by DFT, while approximating the Bloch states as plane waves with free electron dispersion.
        Thus, the supply function can be computed by integrating the density of occupied states over the slice of the reciprocal space which corresponds to a certain normal energy:
        \begin{equation}
          N(E_{z}, T) = \int_{E_{z}}^{\infty} \dif{E} \frac{f_{\text{FD}}(E, T) \rho(E)}{\sqrt{8 m E}} \,. \label{eq:supply}
        \end{equation}
        In the above formula, $f_{\text{FD}}$ is the Fermi--Dirac occupation function, $\rho$ is the density of states and $E$ is the total electron kinetic energy.
        Additionally, although we do not demand zero temperature for the occupation function, the minute perturbative effect of the electron relaxation at non-zero temperatures on the potential is neglected.

        Second, we step away from the simplified model of a jellium metal with a perfectly flat surface.
        Instead we model the metal surface in density functional theory as an atomic system for the computation of the potential barrier.
        However, the electron momentum components parallel to the surface were approximated as having no effect on the transmission probability, which  is exact only under the assumption of a flat surface.
        This approximation is necessary in order to use equation~\eqref{eq:MG-J}.
        Otherwise, computing the differential current density for each electron kinetic energy would have required integrating the transmission probability over the isoenergy surface, which is too computationally demanding.
        \footnote{%
          The calculation replacing equation~\protect\eqref{eq:MG-J} would then take the form
          \begin{equation}
            J = \frac{e}{\sqrt{2 m}} \protect\int_{0}^{\infty} \dif{E} \sqrt{E} f_{\text{FD}}(E) \rho(E) \protect\left\protect\langle D(E, \theta, \phi) \protect\right\protect\rangle \,,
          \end{equation}
          where
          \begin{equation}
            \protect\left\protect\langle D(E, \theta, \phi) \protect\right\protect\rangle = \frac{1}{2 \PI} \protect\int_{0}^{2 \PI} \dif{\phi} \protect\int_{0}^{\PI / 2} \dif{\theta} \sin\theta \cos\theta D(E, \theta, \phi)
          \end{equation}
          is a weighted average of the transmission probability over the upper hemisphere of the isoenergy surface and  $\theta$ and $\phi$ are the polar and azimuthal angles in $\vec{k}$-space, respectively.
          Equation~\protect\eqref{eq:MG-J} with the supply function from equation~\protect\eqref{eq:supply} follows from this when assuming that $D(E, \theta, \phi) = D(E_{z})$, where $E_{z} = E \cos^2\theta$.
        }

        Finally, in order to calculate the transmission probabilities, we did not use the JWKB approximation \cite{landau} as in the classical theory \cite{MurphyG}, or perturbative methods as in recent ab-initio approaches \cite{lepetit2017electronic}. 
        We computed them numerically by using quantum transport theory, as described in section \ref{sec:methods-QT}. 

      \subsubsection{Image potential} \label{sec:theory-changes-image}
        Since the potential obtained directly from DFT calculations can be trusted at the level allowed by the used functional, it is necessary to assess the reliability of the selected functional in all regions of interest, which in our case are bulk, surface and vacuum.
         Although the Perdew--Burke--Ernzerhof (PBE) functional~\cite{perdew1996generalized, *Perdew1996err} used in this work is accurate in and near the metal, it fails to reproduce the long-range exchange-correlation potential in the vacuum~\cite{Eguiluz1992}.
        The true exchange-correlation potential in the vacuum over a smooth metal surface is known to converge to the image potential~\cite{Bardeen1940, Harbola1993, Lang1973}
        \begin{equation}
          V_{\text{xc}}^{\text{im}}(z) = \frac{e^2}{4 \PI \epsilon_{0}} \frac{1}{4 (z - z_{\text{im}})} \,,
        \end{equation}
        where $z_{\text{im}}$ is the location of the electrical surface, also known as the image plane.
        The exchange-correlation potential computed using the PBE functional decays exponentially into the vacuum, which makes the computed potential barriers too high.

        In order to improve the description of the potential barrier, the image potential was externally introduced as an approximation of the true exchange-correlation potential in the vacuum region.
        The image plane was assumed as being flat and parallel to the $x$-$y$ plane.
        Its location is known to be the $z$ coordinate of the center of mass of the charge layer induced by the applied electric field~\cite{Chulkov1997, Chulkov1999, Serena1986}.

        Having computed the exchange-correlation potential $V_{\text{xc}}^{\text{PBE}}$ in the vacuum region with DFT, the total potential was modified as follows to take into account the image potential.
        A smooth transition
        \begin{equation}
          V_{\text{xc}} = f(x) V_{\text{xc}}^{\text{PBE}} + [1 - f(x)] V_{\text{xc}}^{\text{im}}
        \end{equation}
        from the exchange-correlation potential computed using the PBE functional to the image potential was introduced,
        where
        \begin{equation} \label{eq:imtrans}
          f(x) =
          \begin{cases}
            1.0 \,, & x \leq 0 \\
            \E^{- x / \lambda_{x}} + (1 - \E^{- x / \lambda_{x}}) \E^{-x} \,, & x > 0
          \end{cases}
        \end{equation}
        is the a weighting function and $x = (d - d_{0}) / \lambda_{d}$ is a rescaling of the distance $d = \max(r, z') - z'_{\text{im}}$.
        $r$ is the distance of the point for which the potential is being determined from the nearest atom, $z'$ is the distance of the point from the image plane and $z'_{\text{im}}$ is the distance of the image plane from the topmost layer of atoms in the metal, not counting the defect atoms.
        The form of the smoothed Heaviside weighting function $f(x)$ ensures that the resulting exchange-correlation potential is continuous and smooth.
        The $d_{0}$ parameter determines how far from the image plane the transition begins, $\lambda_{d}$ determines how long the transition distance is and $\lambda_{x}$ determines how abrupt the beginning of the transition is.

        The computation of the center of mass was done by splitting the metal slab into two parts in the $z$ direction so that the total charge on each side was that predicted by Gauss' law for the surface charge, and then calculating the quotient of the excess dipole moment of each side and the corresponding surface charge.
        The excess dipole moment was determined by subtracting the dipole moment of the same surface with no applied electric field from the dipole moment of the surface being studied.

  \section{Methods} \label{sec:methods}
  
    \subsection{Simulation setup} \label{sec:methods-setup}
    
\begin{figure}
        \includegraphics[height=1in]{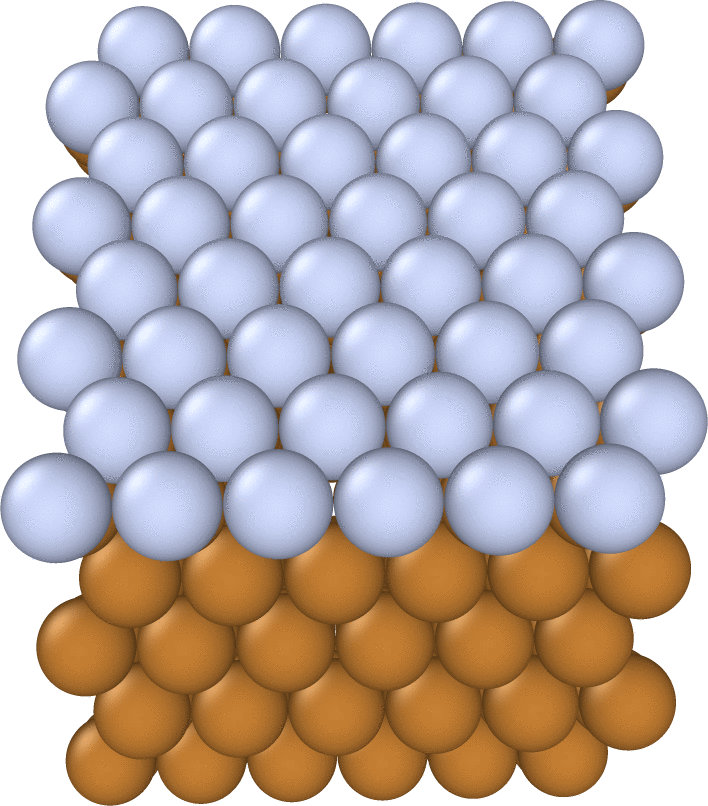}
        \hspace{0.5cm}
        \includegraphics[height=1in]{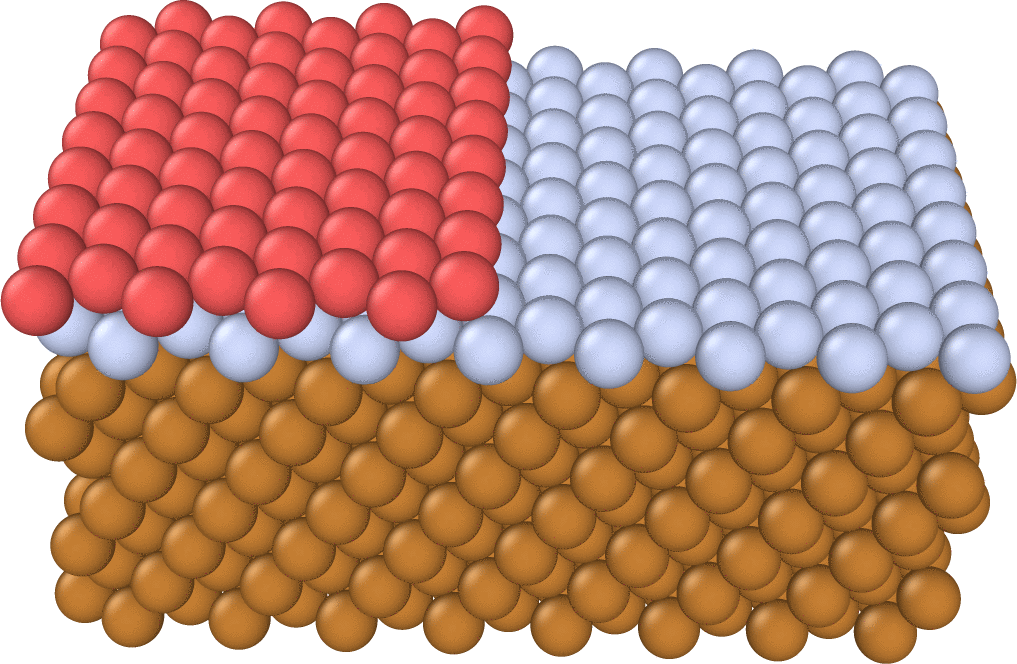}
        \\[0.5cm]
        \includegraphics[height=1in]{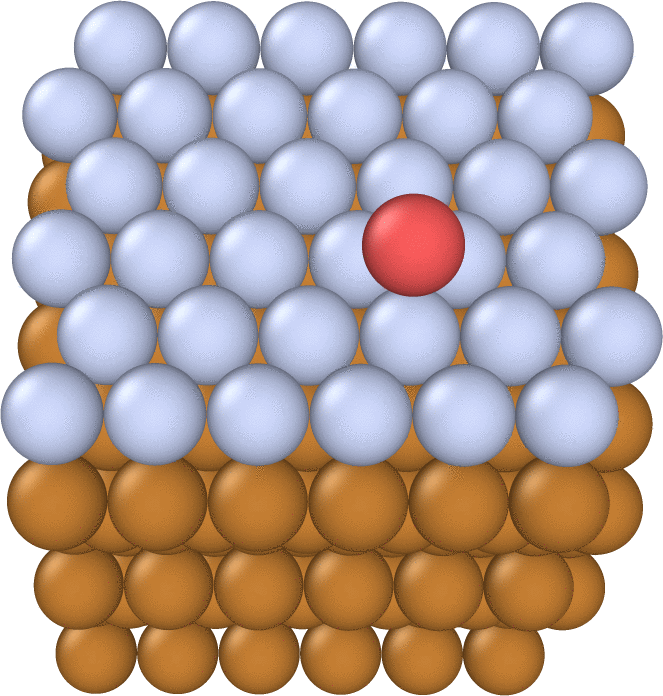}
        \hspace{0.5cm}
        \includegraphics[height=1in]{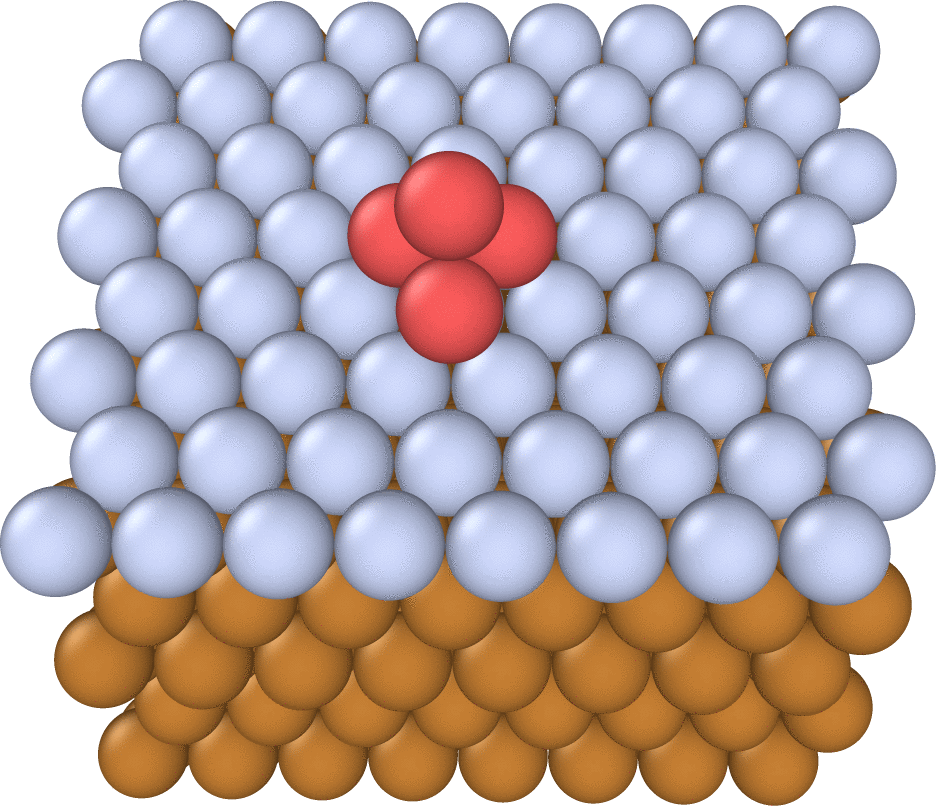}
        \caption{%
          (Color online)
          The studied surface defect types.
          Substrate atoms are orange (bottom layers), surface atoms are blue (top full layer) and atoms added by the defects are red (incomplete layers).
          From top left to bottom right: clean surface, step defect, adatom defect, pyramid defect.
          The extra atoms of the step defect are in FCC sites.
          The adatom is in an FCC site.
          The lower level of the pyramid is in FCC sites.
          The clean surface and step defect images have periodic images added to make the image clearer.
          }
        \label{fig:defect-types}
      \end{figure}

    For this study we selected to to simulate the (111) orientation of a Cu surface, since the expected effect on the work function from atomic defects on this surface is the strongest compared to other orientations.
      Three different types of surface defects were studied along with the clean surface, all shown in figure~\ref{fig:defect-types}.

      The first defect type was a step defect consisting of an additional layer of FCC atoms covering half of the surface.
      To exclude any finite size effect of the step due to periodic boundaries in lateral directions, the length of the cell perpendicular to the step was varied.
      The second defect type was a single adatom in an FCC lattice site on the surface.
      With periodic boundary conditions this resulted in a surface with an adatom density given by the inverse of the area of the supercell.
      The surface was kept approximately square and its size was varied.
      The third defect was a two layer triangular pyramid with the lower layer atoms in FCC lattice sites.
      The system otherwise resembled the adatom defect system.
      The surface was kept approximately square and its size was varied.

      We carried out all DFT calculations using the version 5.3.5 of the Vienna Ab initio Simulation Package (VASP)~\cite{kresse1993ab, Kresse1994, kresse1996efficiency, kresse1996efficient} with the Projector Augmented-Wave method (PAW)~\cite{Bloechl1994b, Kresse1999}.
      The used pseudopotentials were from the version 54 dataset supplied with VASP.
      The used pseudopotential for copper treated the eleven 3d and 4s electrons as valence electrons and the rest of the electrons as a frozen core.
      \subsubsection{Calculation details} \label{sec:methods-setup-details}
        All calculations used the Perdew--Burke--Ernzerhof (PBE) exchange-correlation functional~\cite{perdew1996generalized, *Perdew1996err}.
        The electronic convergence criteria was set to a change of less than \SI{e-7}{\electronvolt} in the total energy and the Kohn--Sham eigenvalues.
        The convergence criterion for the ionic relaxation was set to \SI{e-6}{\electronvolt} in the total energy, which resulted in remaining forces of order \SI{e-3}{\electronvolt\per\angstrom}.

        The convergence of the plane wave energy cutoff, $\vec{k}$-point sampling, smearing method and parameter and lattice constant was tested using a bulk FCC copper system with one atom in the simulation cell.
        The chosen parameters were $E_{\text{cut}} = \SI{800}{\electronvolt}$ for the energy cutoff and $ka \geq \SI{45}{\angstrom}$ for the $\vec{k}$-point sampling, where $a$ is the simulation cell size and $k$ the number of $\vec{k}$-samples in each direction.
        A value $\sigma = \SI{0.2}{\electronvolt}$ was chosen for the first order Methfessel--Paxton~\cite{methfessel1989high} smearing parameter, and $a_{0} = \SI{3.635}{\angstrom}$ for the lattice constant.
        The energy cutoff and $\vec{k}$-point sampling were chosen so that the estimated error from each was less than \SI{1}{\milli\electronvolt}, the smearing parameter was chosen to give a small energy difference to using the tetrahedron smearing method with Blöchl's corrections~\cite{Bloechl1994a}, and the lattice constant was determined with a resolution of \SI{1}{\milli\angstrom}.

        The unit cell for the surface calculations was chosen to be the minimal orthorhombic FCC unit cell with the $(111)$ direction along the $z$ axis and the $x$ direction being along the shorter cell edge.
        The unit cell contains three layers of two atoms.
        The lowest layer has the atoms in the corner and in the middle of the cell, and the atoms in the other layers are shifted by a third of the cell length in the $y$ direction each.
        The ionic positions were relaxed in all surface calculations with the bottom \num{2} layers being fixed to the bulk positions.

        We also performed convergence tests with regard to the number of slab layers and the thickness of the vacuum layer, using a system consisting of a single unit cell in the $x$-$y$ plane with a variable number of slab layers and variable vacuum thickness.
        Additionally, we studied the convergence of the work function and the interlayer spacing of the slab.
        In these tests, we found that the inter-layer spacing converges for a slab thickness of \num{8} atomic layers.
        A vacuum layer of approximately 30 Å was sufficient to achieve convergence of the work function within 10 meV uncertainty while leaving enough vacuum space for the defect structures.

        The convergence with regard to the size of the supercell parallel to the surface was done separately for each defect type.
        The used criterion for convergence was the effect of the defect on the work function being approximately inversely proportional to the size of the surface.
        The criterion effectively means that the area furthest from the defect is equivalent to the clean surface.
        The resulting surface sizes were $1 \times 8$ unit cells for the step defect and $4 \times 2$ unit cells for the adatom and pyramid defects.
    \subsection{Potential barriers} \label{sec:methods-barrier}
      The potential barrier of interest for the tunneling of electrons into the vacuum is the one found at the surface of the slab.
      The potential oscillations inside the slab due to the atoms are far below the Fermi level and not of interest in the current work.
      The total local potential was taken from the DFT calculations described above merged with the image potential as described in section~\ref{sec:theory-changes-image}.

\begin{figure}
        \includegraphics{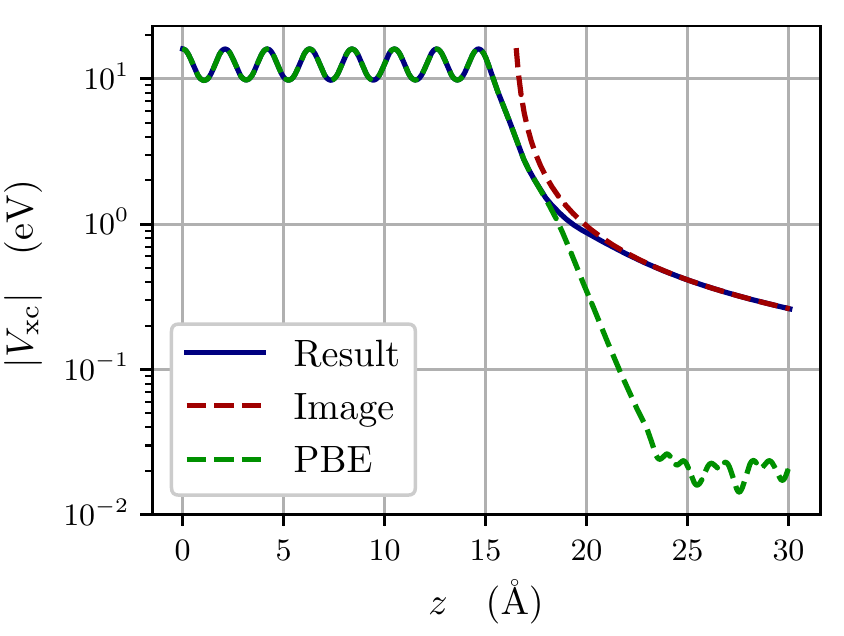}
        \caption{%
          (Color online)
          The transition between exchange-correlation potentials for the clean surface with an applied electric field of \SI{2}{\giga\volt\per\meter}.
          The PBE exchange-correlation potential, the image potential and the resulting exchange-correlation potential are shown.
          The noise in the vacuum region for the PBE potential is a result of a small ($\approx \SI{e-10}{\per\cubic\angstrom}$) amount of electrons in the vacuum due to the numerical approximations in the DFT calculation.
          }
        \label{fig:image-xc}
      \end{figure}

      The parameters for the transition from the PBE potential to the image potential were found by optimizing them for the clean surface, ensuring that the potentials merge smoothly near the region where the distance between them is the shortest (see Fig. \ref{fig:image-xc}).
      This optimization resulted in choosing the following values for the parameters used in expression \eqref{eq:imtrans}: $d_{0} = \SI{1.5}{\angstrom}$, $\lambda_{d} = \SI{1.0}{\angstrom}$ and $\lambda_{x} = \SI{0.75}{\angstrom}$.

      The center of mass of the excess charge was found to be \SI{1.56 \pm 0.08}{\angstrom} above the top atomic plane in the slab with the clean surface.
      This is slightly further out than the result for jellium slabs of $\approx \SI{1.13}{\angstrom}$~\cite{Chulkov1999}.

      The exchange-correlation potential $V_{\text{xc}}^{\text{PBE}}$ was computed as the difference between the total local potential and the Hartree potential, which are outputs of VASP.
      Due to unphysical noise in the exchange-correlation potential approximately \SIrange{5}{8}{\angstrom} above the surface layer, $V_{\text{xc}}^{\text{PBE}}$ was numerically smoothed.
      The smoothing removed the noise while only negligibly effecting the form of the potential.
      We performed the smoothing by applying to each grid line in the $z$ direction a seventh order, \num{71}-point ($\approx \SI{1.75}{\angstrom}$) Savitzky--Golay filter.
      After that, we applied two uniform filters, the first with a $5 \times 5 \times 5$ grid point stencil and the second with a $3 \times 3  \times 3$ grid point stencil.

      Figure \ref{fig:image-xc} demonstrates the shape of the modified exchange-correlation potential in a solid line. 
      In the same figure, the original PBE and image potentials are shown in dashed lines with green and red colors, respectively. 

    \subsection{Quantum transport} \label{sec:methods-QT}
      This section describes the steps necessary to calculate the field emission current densities from the studied surfaces of interest (see Fig. \ref{fig:defect-types}), using the supply function and the potential grid from the DFT calculations, corrected to include the image potential.
      
      The numerical quantum transport calculations were carried out using the Kwant code~\cite{Groth2014} (version 1.3.2), which implements the computation of transport properties for systems described by tight-binding Hamiltonians.
      The latter result from applying the finite difference method on the single-particle Schr\"odinger equation, assuming a certain potential distribution $V$, which is pre-calculated by DFT.
      The resulting tight-binding Hamiltonian includes only onsite terms $V + 2 \sum_{i} T_{i}$ and nearest-neighbor interactions $-T_{i}$, where $T_{i} = \hbar^2 / (2 m a_{i}^{2})$ is the hopping energy for the grid spacing $a_{i}$ in the direction $i$.
      In the tight-binding formalism, semi-infinite leads with uniform constant potential are defined for the metal and vacuum regions perpendicular to the surface.     
      Then the scattering matrix of the transport modes in the leads is calculated by the wave function matching method. 

      Since the quantum transport calculations were done in the jellium approximation, it was necessary to define the constant potential level inside the metal.
      The output from DFT only gives the potential barrier in relation to the Fermi level, so the constant potential in the metallic lead had to be determined otherwise.
      Here we adopted the standard practice of choosing it as the bottom of the valence band \cite{forbes2009use,jensen2007advances}.
      However, we found that the if the zero potential is sufficiently deep (several eV below $E_F$), its position does not affect significantly the resulting transmission coefficients.
      The density of states in all calculations was nonzero only in a continuous area around the Fermi level.
      Hence, our choice is determined by the lowest energy valence electron state, which has zero kinetic energy in the used electron model.

\begin{figure}
          \includegraphics{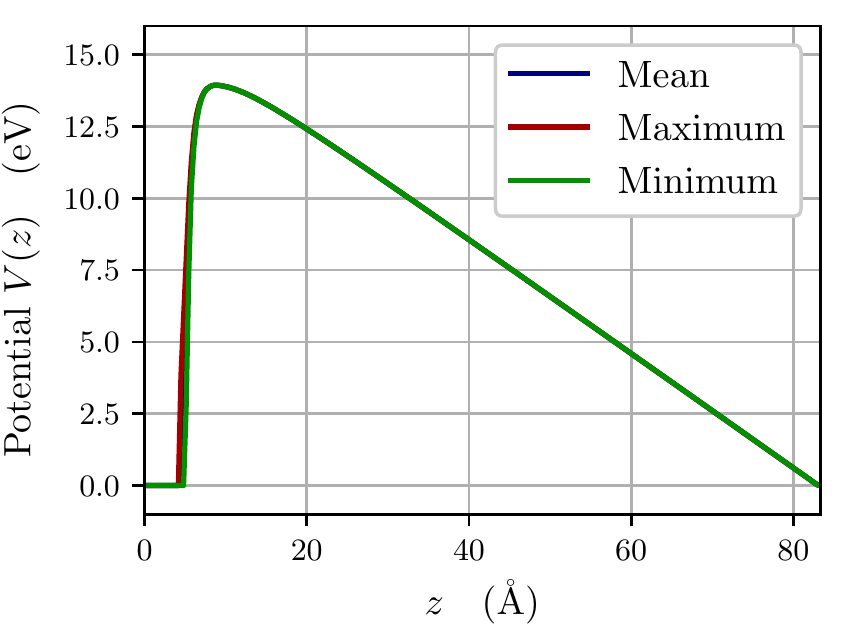}
          \caption{%
            (Color online)
            The potential outputted by VASP after processing.
            The mean, maximum and minimum of the potential is computed for each plane.
            The zero energy is fixed so that the Fermi level has the correct height below the top of the potential barrier.
            The system has a clean surface and an applied field of \SI{2}{\giga\volt\per\meter}.
            The minimum and maximum values are nearly identical in most of the plot.
            }
          \label{fig:potential-processed}
        \end{figure}

      After adding the image potential in the vacuum region, the total potential had to be processed in order to be used in the quantum transport calculations.
      Details of this procedure are given in appendix~\ref{app:methods-QT-process}.
      The shape of the processed potential is shown in figure \ref{fig:potential-processed}.

        We used the processed potential as the scattering section of the system in the quantum transport calculations.
        Identical leads were attached to the metal and the vacuum.
        The leads were at the zero potential and had the same translational symmetry as the potential grid, so the wave functions in the leads were plane waves.
        While this is correct inside the metal in the jellium model, the potential far away in the vacuum should decay linearly due to the electric field.
        However, this difference in the potential is far below the Fermi level, so its effect on the calculations was negligible.

        Using the planar surface approximation for the supply function means that the momenta of the incoming waves parallel to the surface are approximated as irrelevant for the transmission probability.
        They were therefore set to zero, making the only relevant incoming mode the one with zero transverse energy.
        The transmission probability was computed as the sum of the transmission probabilites from the lowest mode in the metal lead to all modes in the vacuum lead.
        The single-mode transmission probabilities were computed from the scattering matrix, the primary output of the Kwant calculation.

        The transmission probabilities were computed for energies from \SI{3}{\electronvolt} below the Fermi level to \SI{5}{\electronvolt} above the Fermi level with a resolution of \SI{0.1}{\electronvolt}.
        The resolution was changed to \SI{0.025}{\electronvolt} for \SI{0.1}{\electronvolt} to either side of the Fermi level due to the importance of this area for the field emission current.

        After computing the transmission probabilities, the differential current densities were computed as the product of the transmission probability and the supply function.
        The total current density from the studied surface was then computed by integrating the differential current density over the normal energy, as in eq. \eqref{eq:MG-J}.
        The numerical integration was done using Simpson's rule.
  \section{Results} \label{sec:results}
    \subsection{Work functions} \label{sec:results-workfunctions}
\begin{figure}
        \includegraphics{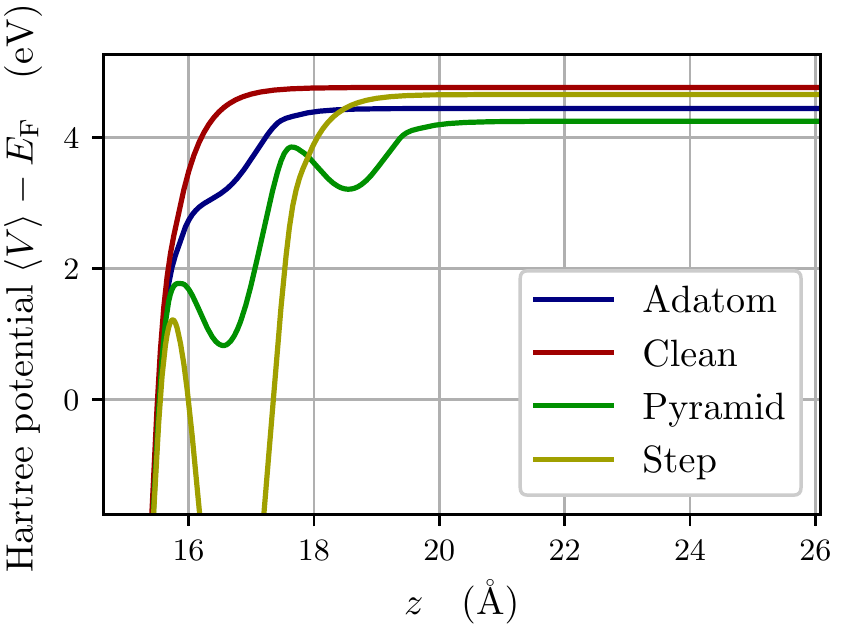}
        \caption{%
          (Color online)
          Hartree potentials for the systems without applied electric fields.
          The zero potential is the Fermi level.
          The work functions can be read off as the potential value at the right edge of the figure.
          }
        \label{fig:workfunctions}
      \end{figure}

      We computed the work function for each surface as the difference between the converged Hartree potential in the vacuum and the Fermi level using the calculations without applied electric field.
      The Hartree potential was fully converged to the vacuum value in each calculation, as can be seen in figure~\ref{fig:workfunctions}.

      The work function values for the converged surface sizes are shown in table~\ref{tab:workfunctions}.
      One can see that, as expected, the work functions are lower for the defect systems than for the clean surface.
      The effect on the work function is largest for the pyramid defect and smallest for the step defect, meaning that the defects which cause a rougher surface have smaller work functions.
      This is expected due to the effect of Smoluchowski smoothing~\cite{Smoluchowski1941}.

      \begin{table}
        \begin{ruledtabular}
          \begin{tabular}{lddd}
            System & \tableheader{1}{Work function $\phi \units{\si{\electronvolt}}$} & \tableheader{1}{$\Del{\phi} \units{\si{\electronvolt}}$} & \tableheader{1}{$\Del{\phi} \units{\si{\percent}}$} \\
            \hline \\[-1.5ex] 
            Clean surface  & 4.76 & 0.00  & 0.0  \\
            Step defect    & 4.66 & -0.10 & -2.1  \\
            Adatom defect  & 4.44 & -0.32 & -6.7  \\
            Pyramid defect & 4.25 & -0.51 & -10.7
          \end{tabular}
        \end{ruledtabular}
        \caption{%
          The work functions determined for the different surface defects with converged surface sizes.
          $\Del{\phi}$ is the difference to the clean surface work function.
          }
        \label{tab:workfunctions}
      \end{table}

      The work function of the clean surface is expected to be lower than the experimental value of \SI{4.94}{\electronvolt}~\cite{Gartland1972}, as the PBE functional underestimates work functions by \SI{0.3}{\electronvolt} on average~\cite{DeWaele2016}.
      The computed value is in good agreement with the literature value for the PBE functional, \SI{4.78}{\electronvolt}~\cite{DaSilva2006}.
      The reduction of the work function by \SI{6.7}{\percent} for the adatom defect is in good agreement with the calculations of Djurabekova et al.~\cite{Djurabekova2013}, which yielded a reduction of \SI{5.9}{\percent} for the copper $(100)$ surface using the same criteria for the size of the surface.

      The validity of comparing the computed work functions of different systems is discussed in section~\ref{sec:discussion-work}.
    
    \subsection{Transmission probabilities and emission currents} \label{sec:results-probcurr}
\begin{figure}
        \includegraphics{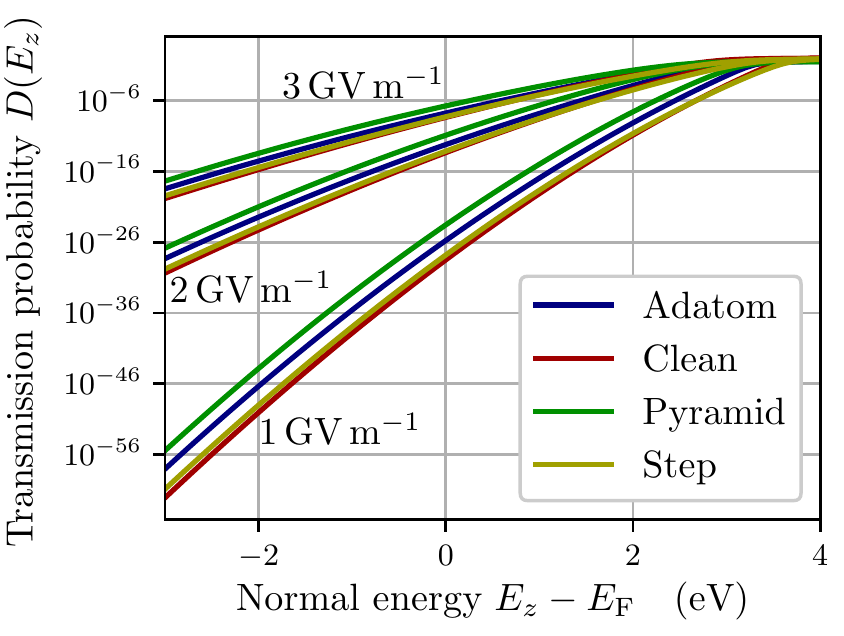}
        \caption{%
          (Color online)
          Transmission probabilities for the different surface defects as a function of the normal energy.
          The plot for the clean surface and step defect are almost on top of each other for $F > \SI{1}{\giga\electronvolt\per\meter}$.
          }
        \label{fig:probabilities}
      \end{figure}

      Figure~\ref{fig:probabilities} shows the computed transmission probabilities as a function of the normal energy.
      One can see that the transmission probability is higher for a system with a smaller work function or a larger applied electric field.
      The order of the transmission probabilities for the different defect types is the same for all applied electric fields, and is the same as in table~\ref{tab:workfunctions}.

      All transmission probability curves are qualitatively similar to those for the Schottky--Nordheim barrier.
      Notably, all of the curves are almost parallel near the Fermi level, with differences in their slope appearing only at higher energies near the top of the barrier, which are irrelevant for field emission due to the Fermi--Dirac statistics causing the supply function to vanish.
      The curves are steeper for the lower applied fields, because the width of the barrier increases faster as the energy decreases.

      The curvature of the plots near the Fermi level is small, which justifies the use of the standard Fowler--Nordheim equation for finding the field enhancement factors, as its derivation is based on a linear approximation of the transmission probability near the Fermi level.

\begin{figure}
        \includegraphics{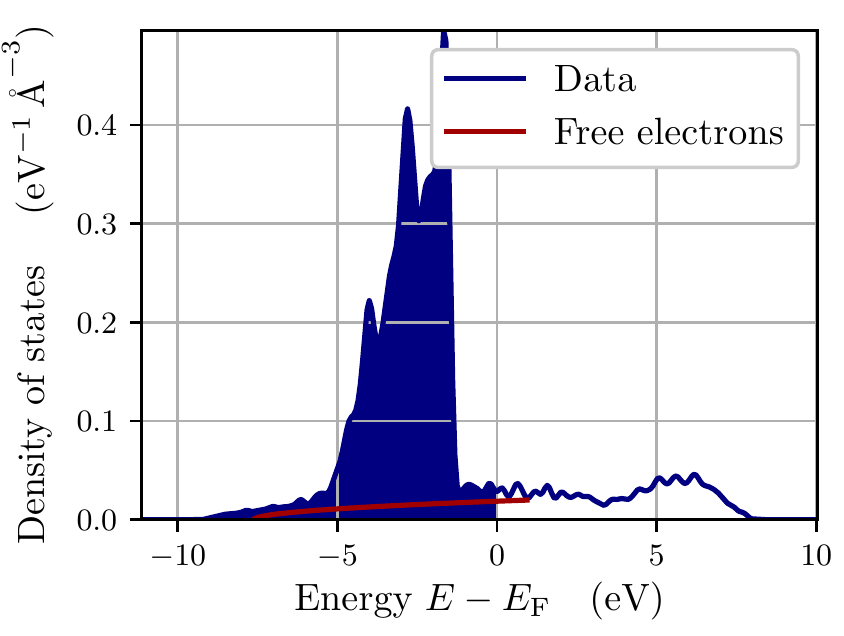}
        \caption{%
          (Color online)
          The density of states computed for the system with a clean surface and no applied electric field.
          The densities of states for the other systems showed no significant differences.
          The free electron density of states corresponding to the same electron density in the slab is shown for comparison.
          }
        \label{fig:dos}
      \end{figure}

      The computed densities of states for the different systems differed from each other only negligibly.
      An example can be seen in figure~\ref{fig:dos}.
      They closely resembled the density of states of bulk copper, with a small background from the 4s electrons and a peak approximately \SI{1.5}{\electronvolt} below the Fermi level from the 3d electrons.
      Since the energies with large densities of states are significantly below the Fermi level, they are not of significant importance for field emission.
      Neglecting the peak, one could approximate the density of states by the free electron parabola, which would result in the introduction of a constant correction factor to the emission current.

\begin{figure}
        \includegraphics{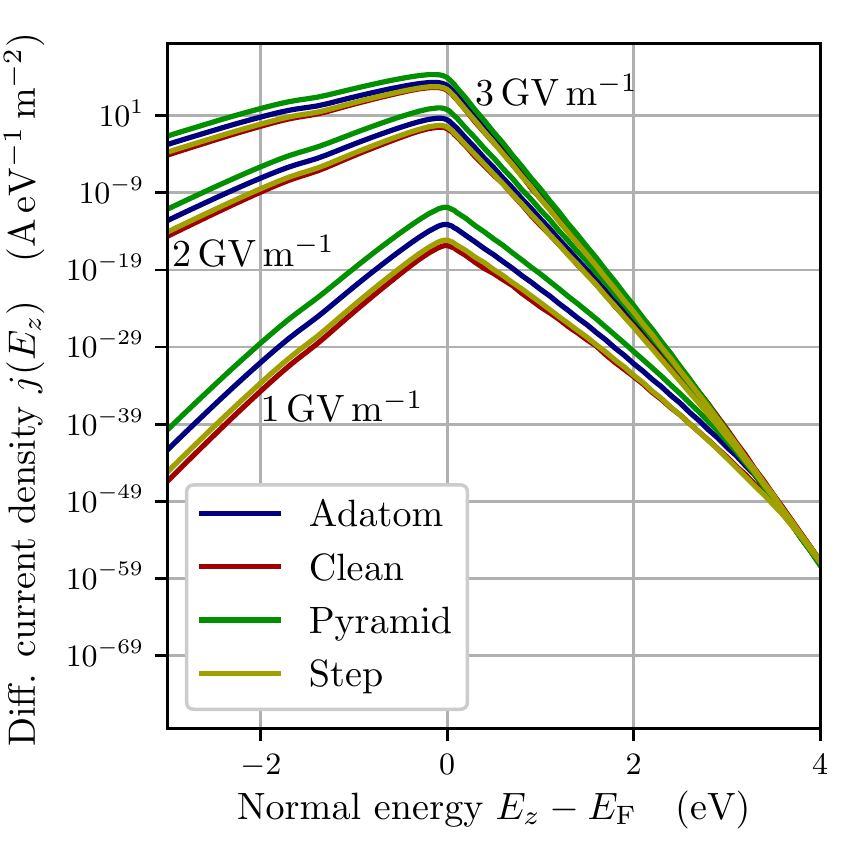}
        \caption{%
          (Color online)
          The differential current densities for the different surface defects as a function of the normal energy at room temperature.
          At zero temperature the differential current density vanishes abruptly at the Fermi level but is almost exactly the same for $E_{z} < E_{\text{F}} - \SI{0.1}{\electronvolt}$.
          The plot for the clean surface and step defect are almost on top of each other for $F > \SI{1}{\giga\electronvolt\per\meter}$.
          }
        \label{fig:currents}
      \end{figure}

      The differential current densities $j(E_{z})$, the emission current density per unit normal energy from electrons with the normal energy $E_{z}$, is shown at room temperature as a function of the normal energy in figure~\ref{fig:currents}.
      The temperature determines the slope of the decrease above the Fermi level, which becomes infinite for zero temperature. 
      Thus, the finite temperature contributes to the emission current from energies above the Fermi level.

      The qualitative comparison of the curves for different systems similar to the one for the transmission probabilities, since the density of states which also affects $j(E_{z})$ is almost the same for all systems.
      However, the irregular shape of the computed density of states has a visible effect on the $j(E_{z})$ curves, appearing as a slight increase of $j(E_{z})$ at $E_z \approx$ \SI{-1.5}{\electronvolt}.

      The integrated total currents are listed in table~\ref{tab:currents}.
      One can see that the current density for the Schottky--Nordheim barrier is larger than the current density computed for the clean surface by a factor of approximately \numrange{2}{5} for all computed electric field strengths.
      Overall, one can see that the differences between the systems become smaller as the electric field increases.
      The increase of the emission current caused by the defects is large, about \num{3} orders of magnitude for the single adatom and \num{6} for the pyramid defect with the smallest electric field.

      \begin{table}
        \begin{ruledtabular}
          \begin{tabular}{l|dldldl}
            \diagbox{System}{Field} & \tableheader{2}{\SI{1}{\giga\volt\per\meter}} & \tableheader{2}{\SI{2}{\giga\volt\per\meter}} & \tableheader{2}{\SI{3}{\giga\volt\per\meter}} \\
            \hline \\[-1.5ex] 
            S.--N. barrier & 3.96 & $\times 10^{-17}$ & 2.19 & $\times 10^{-1}$ & 4.69 & $\times 10^{4}$ \\
            Clean surface  & 8.75 & $\times 10^{-18}$ & 9.23 & $\times 10^{-2}$ & 1.76 & $\times 10^{4}$ \\
            Step defect    & 1.33 & $\times 10^{-16}$ & 1.52 & $\times 10^{-1}$ & 1.84 & $\times 10^{4}$ \\
            Adatom defect  & 1.15 & $\times 10^{-14}$ & 1.21 &                  & 7.98 & $\times 10^{4}$ \\
            Pyramid defect & 9.53 & $\times 10^{-12}$ & 2.66 & $\times 10^{1}$  & 7.13 & $\times 10^{5}$
          \end{tabular}
        \end{ruledtabular}
        \caption{%
          Field emission electron currents for the different systems and electric fields at zero temperature in \si{\ampere\per\square\meter}.
          The results for the Schottky--Nordheim barrier are shown for comparison.
          }
        \label{tab:currents}
      \end{table}
    \subsection{Fowler--Nordheim plot analysis} \label{sec:results-FNanalysis}
    In order to assess the effects of the various surface irregularities on the parameters usually extracted by field emission measurements, we shall perform Fowler--Nordheim plot analysis on the calculated current densities.
    For this purpose, we shall linearize eq. \eqref{eq:FN_comp} by approximating the parameter $\nu(F)$ as a linear function of the applied electric field, $\nu = \nu_{0} + \nu_{1} F$ and the parameter $\tau$ as a constant $\tau_{0}$.
      The parameters $\nu_{0}$ and $\nu_{1}$ were computed from least squares fits to the computed emission currents, and it was found that $\nu_{0} \approx 1$ for all systems while $\nu_{1}$ varied.
      The linear approximation was justified post hoc, as it leads to linear Fowler--Nordheim plots, which were observed.
      The resulting linearized form of the Fowler--Nordheim plots is
      \begin{equation}
        \underbrace{\ln\left( \frac{J}{F^2} \frac{\phi}{a} \right)}_{Y} = - \frac{1}{\beta} \underbrace{\frac{\nu_{0} b \phi^{3/2}}{F}}_{X} - \underbrace{\frac{\nu_{1} b \phi^{3/2}}{\beta} + \ln\left( \lambda \frac{\beta^2}{\tau_{0}^2} \right)}_{\text{constant}} \,. \label{eq:FN-lin}
      \end{equation}

\begin{figure}
        \includegraphics{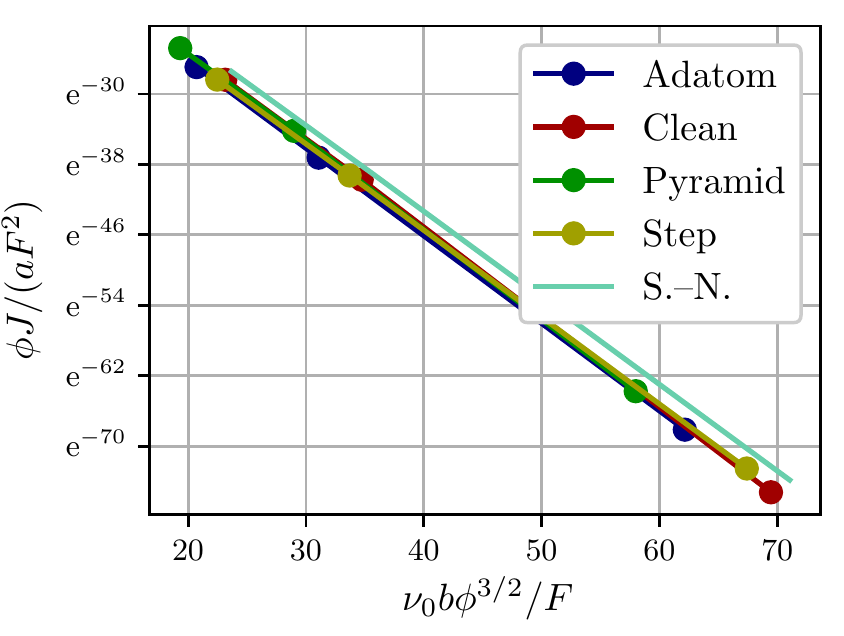}
        \caption{%
          (Color online)
          Fowler--Nordheim plots for the different surface defects at zero temperature.
          The Schottky--Nordheim barrier result with no field enhancement is shown for comparison.
          In the chosen linearization the field enhancement factor is the reciprocal of the negative slope.
          }
        \label{fig:FN-plot}
      \end{figure}

      The field emission currents computed for the different defect type and applied electric fields are plotted in a Fowler--Nordheim type plot of the form \eqref{eq:FN-lin} in figure~\ref{fig:FN-plot}.
      One can see that the plots are linear, with slopes very similar to the F-N equation (S-N barrier), which is unity.
      We shall now define the effective field enhancement factor (or slope correction factor) $\beta$ for each system, as the inverse of the slope of each curve.
      We call it effective field enhancement because it accounts for all the possible effects that induce a change in the slope of the F-N plot, equivalent to an actual electrostatic field enhancement of the same magnitude.

     In figure \ref{fig:FN-plot}, curves have similar slopes that deviate less than 2\% from unity. 
      Thus, the effective field enhancement factors induced by the simulated defects are insignificant.
      This was also verified using linear fits to the plots, which show that $\beta$ is unity to the accuracy allowed by this work.
      This means that the increase in the emitted currents due to the surface defects can only be a result of the decrease of the local work function.

      The remaining difference between the computed plots and the reference plot for the S--N barrier is simply an approximately constant factor, which is smaller for the studied systems than for the reference system.
      This means that the technically complete Fowler--Nordheim equation \eqref{eq:FN_comp} is sufficient to describe the field emission from the studied surfaces, given that correct values for the work function $\phi$ and the constant pre-factor $\lambda$ have been obtained.

	  However, in standard experimental circumstances, it is impractical to obtain an accurate value of $\phi$ for defective surfaces, and a conventional clean-surface value is usually assumed.
	  In this case, the change in the slope of the F-N plot due to the different $\phi$ is interpreted as field enhancement.
      The apparent field enhancement factor $\beta_{\text{app}}$, i.e. the $\beta$ extracted from the F-N plot when a certain value $\phi_{\text{a}}$ is assumed for the work function, can be deduced from equation~\eqref{eq:FN-lin} (neglecting small differences in $\nu_0$) as
      \begin{equation}
        \beta_{\text{app}} = \left( \frac{\phi_{\text{a}}}{\phi} \right)^{3/2} \beta \,,
      \end{equation}
      where $\phi$ is the real work function and and $\beta$ the effective field enhancement factor of the surface (unity in our calculations).
      The ratio of the assumed and real work functions is greater than unity because of the surface defects.

\begin{figure}
        \includegraphics{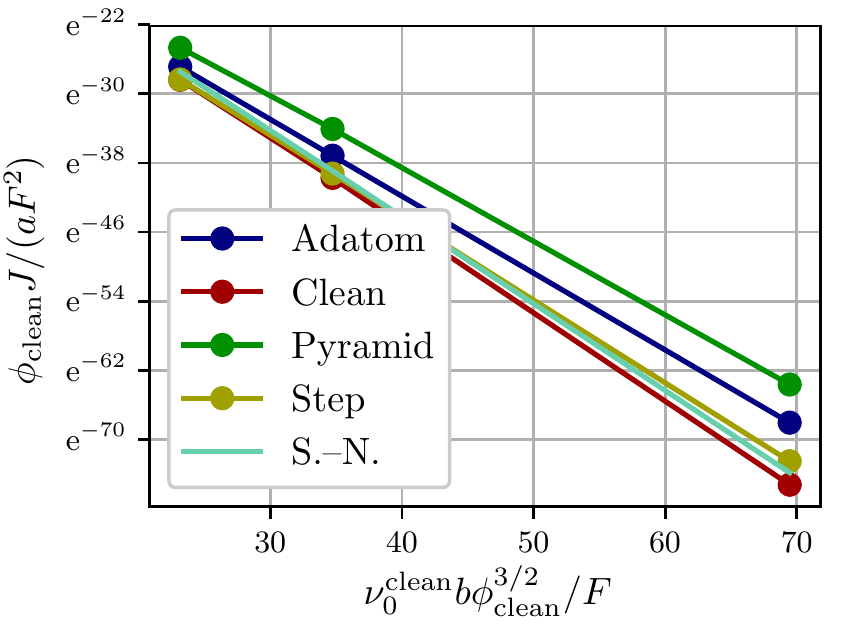}
        \caption{%
          (Color online)
          Fowler--Nordheim plots for the different surface defects at zero temperature.
          The work function and $\nu_{0}$ parameter are set to those of the clean surface in all plots.
          The Schottky--Nordheim barrier result with no field enhancement is shown for comparison.
          In the chosen linearization the field enhancement factor is the reciprocal of the negative slope.
          }
        \label{fig:FN-plot-cleanwork}
      \end{figure}

      Figure \ref{fig:FN-plot-cleanwork} shows the same plots as figure \ref{fig:FN-plot}, with the difference that the work function and barrier enhancement parameter are assumed to be those of the clean surface for all surface defects.
      One can see that the slopes of the plots are now different due to the scaling of the abscissa with the work function.
      The apparent field enhancement factor of the adatom and pyramid defects, as extracted using eq. \eqref{eq:FN-lin} are now approximately \num{1.14} and \num{1.24} respectively.

  \section{Discussion} \label{sec:discussion}
    \subsection{Work function determination} \label{sec:discussion-work}
      Much of the analysis in this work is based on quantifying the effect of a surface defect on the work function of a metal surface.
      The work function as computed here, is effectively a property averaged over the whole cell surface.
      It is therefore clear that the effect of a single defect on the calculated work function depends on the size of the cell, or in other words the coverage density of defects.
      This means that the comparison of the work function changes among different defect types is meaningful only if they all have the same size, i.e. the same surface coverage.
     
     Furthermore, the cell size has to be sufficiently large, so that the defects do not interact with their periodic images. 
     In order to assure this, we adopted a method given in an earlier work by Djurabekova et al~\cite{Djurabekova2013}, which studied the effect of an adatoms and step defects on a copper $(100)$ surface on the work function.
      The change of the work function resulting from a change in the lateral size of the slab is studied and convergence is reached when the change in the work function due to the defect is inversely proportional to the surface size, $\Del{\phi} \propto 1 / S$, or equivalently proportional to the coverage.
      This corresponds to a situation where the local changes around the defect do not affect the area near the periodic boundaries and the defects can therefore be considered as isolated.
      For one dimensional defects such as the step, the convergence criterion must be modified to take into account that the coverage depends only on the length of the supercell perpendicular to the step, $\Del{\phi} \propto 1 / L_{y}$.
    \subsection{Apparent field enhancement} \label{sec:discussion-enhancement}
    One important result of this paper is that small surface defects, such as the ones simulated here, do not cause any significant change to the slope of the Fowler--Nordheim plot, apart from the one attributed to the modification of the work function. 
    In other words, they do not induce any significant effective field enhancement in the field emission characteristics of a metal surface.
    Therefore, such defects alone, cannot explain the increased field emission currents and apparent field enhancements measured experimentally from clean Cu surfaces~\cite{Kildemo2004}.
      However, the reduction of the work function, which is not taken into account when analyzing F--N plots (a uniform value $\phi=$4.5 eV is usually considered for Cu~\cite{Kildemo2004}), might contribute in the overestimation of the field enhancement. 

     Although these apparent field enhancement factors due to work function decrease are small, they are produced by \si{\angstrom}-scale defects of copper.
      Surface contaminants which lower the work function (see e.g. ref.~\onlinecite{Leung2003}) and larger defects which cause more surface roughness can cause much larger apparent field enhancement factors, which might provide an explanation for at least part of the field enhancement factors reported in experimental works.

      The effect of finding large field enhancement factors due to disregarding the local work function has been studied before~\cite{Chen2012, Qian2012}, where it was found that the field enhancement factors for clean surfaces are approximately \numrange{1}{5} if the local work function is significantly lowered.
      
      \subsection{Limitations of the method}
      The method developed in this work is valuable for calculating field emission characteristics beyond the classical models.
      However, there are certain limitations imposed by the approximations that have been adopted.
      
      First, the electrical surface of the slab, used to define the image plane, was assumed to be flat for all of the studied systems, even those with defects.
      This is due to the image potential only being known as the correct asymptotic exchange-correlation potential in the vacuum for a flat surface.
      This approximation limits the sizes of the surface and the defect structure which can be studied using the method described.
	  A very high defect such as a nanorod would raise the computed electrical surface high above the surface of the surrounding clean surface.
      The effect of the image potential on the potential barrier there would therefore be removed and the potential would be too high.
      Also, using the image potential in the area around the nanorod would be questionable.

	Second, the electric fields accessible by the specific DFT method used here are limit to about 3 GV/m. 
	The standard field range for which field emission measurement from materials with Cu-like work function are taken is 3-7 GV/m \cite{CabreraPRB}. 
	Thus, the highest field achieved in our calculations lies in the lower limit for experimentally detectable field emission currents. 
	However, although the current densities calculated for lower fields (1-2 GV/m) are not experimentally measurable, they are theoretically invaluable for calculating the slope of F--N plots and comparing with classical emission models.

  \section{Conclusions} \label{sec:conclusions}
    We developed a methodology for performing \textit{ab initio} calculations of field emission currents from metal surfaces using density functional theory and quantum transport calculations.
    The method was then used to analyze the effect of atomic-scale defects on copper surfaces on the field emission current and to determine the validity of the Fowler--Nordheim equation for these surfaces.
    The defects were found to decrease the work function, with the decrease being larger for defects with higher roughness.
    The largest local decrease of the work function was found to be \SI{0.51}{\electronvolt}, corresponding to an apparent field enhancement of 1.24 if the flat-surface work function were used.
    The analysis of the field emission currents determined using the developed method showed that the Fowler--Nordheim equation provides a good qualitative description of the field emission from surfaces with atomic-scale defects.
    For a quantitative comparison, the equation must be extended to the technically complete Fowler--Nordheim equation with a multiplicative pre-factor which depends on the specific material and surface configuration and can be calculated using our method.
    Building up on this work, the developed methodology can be used and expanded to calculate field emission characteristics of more complicated small surface structures and defects, or materials that cannot be described with the classical emission models.
    
    \section*{Acknowledgments}
H. Toijala and A. Kyritsakis were supported by the CERN K-contract (No. 47207461). The work of K. Eimre and V. Zadin was supported by the Estonian Research Council grant PUT 1372. F.\;Djurabekova acknowledges gratefully the financial support of the Academy of Finland (Grant No. 269696). We also acknowledge the grants of computer capacity from the Finnish Grid and Cloud Infrastructure (persistent identifier urn:nbn:fi:research-infras-2016072533).

  \appendix

    \section{Details of processing the potential} \label{app:methods-QT-process}
      After adding the image potential in the vacuum region, the potential still had to be processed in order to be used in the quantum transport calculations where the jellium model for the slab is considered.
      First, the potential was cropped to the dipole correction, removing the periodic image of the lower side of the slab.
      Second, most of the slab was cropped from the potential, leaving only \SI{4}{\angstrom} of slab below the top atomic plane of the substrate.
      This was only done to reduce the computational expense of the calculations and has no effect on the results.

      Third, the potential in the vacuum was extended to the zero potential level of the jellium model.
     This was necessary because of the limited vacuum height used in the DFT calculations.
      A least squares fit was made to the potential for each grid line in the $z$ direction within the upper third of the vacuum region.
      The fitted function was of the form
      \begin{equation}
        V(z) = A - eFz - \frac{e^2}{4 \PI \epsilon_{0}} \frac{1}{4(z - z_{\text{im}})} \,,
      \end{equation}
      where $A$ is a fitting parameter, $F$ is the applied electric field and $z_{\text{im}}$ is the location of the image plane.
      Using the applied electric field instead of making it a fitting parameter is justified, as checking the potential output by VASP revealed that the electric field was equal to the applied electric field everywhere in this region within the margin of numerical error.
      Fitting the electric field would have caused slightly different values for neighboring grid lines and therefore discontinuities in the continued potential far from the surface.

      Fourth, the potential grid was undersampled \num{12} times, in order to reduce the required computational resources to feasible levels, while maintaining a sufficient accuracy.
      This was found to increase the computed transmission probability near the Fermi level by approximately \SI{20}{\percent} compared to undersampling the potential only \num{4} times (computing a reference using no undersampling was not feasible).
      The undersampling did not break the translational symmetry of the potential grid.
      The resulting potential grid had a spacing of approximately \SI{0.32}{\angstrom}, allowing plane waves with up to \SI{37.6}{\electronvolt} kinetic energy to be described, while the largest energy required for the calculations was approximately \SI{15}{\electronvolt}.

      Finally, the potential inside the slab was clamped to the zero potential in order to emulate a jellium slab.
      For each grid line in the $z$ direction, the last place where the potential crossed the zero potential before the surface barrier was found.
      If this did not happen for a grid line, the $z$ coordinate of the top atomic plane was chosen.
      The potential in the area below the surface determined above was set to the zero potential.
  \bibliography{bibliography/bibliography}
\end{document}